\documentclass[aps,twocolumn,showpacs,floatfix,amssymb]{revtex4}
\usepackage{graphicx}
\begin{document}

\title{Spin excitations and thermodynamics of the antiferromagnetic Heisenberg model
 on the layered  honeycomb  lattice}
\author{ A.A. Vladimirov$^{a}$, D. Ihle$^{b}$ and  N. M. Plakida$^{a}$ }
 \affiliation{ $^a$Joint Institute for Nuclear Research,
141980 Dubna, Russia}
 \affiliation{$^{b}$ Institut f\"ur Theoretische Physik,
 Universit\"at Leipzig,  D-04109, Leipzig, Germany }

\date{\today}

\begin{abstract}
We present a spin-rotation-invariant Green-function theory for the dynamic spin
susceptibility in the spin-1/2 antiferromagnetic Heisenberg model on a stacked
honeycomb lattice. Employing a generalized mean-field approximation for arbitrary
temperatures, the thermodynamic quantities (two-spin correlation functions, internal energy,
magnetic susceptibility, staggered magnetization, N\'{e}el temperature, correlation length)
and the spin-excitation spectrum are calculated by solving a coupled system of
self-consistency equations for the correlation functions. The temperature dependence
of the magnetic (uniform static) susceptibility is ascribed to antiferromagnetic
short-range order. The N\'{e}el temperature is calculated for arbitrary interlayer
couplings. Our results are in a good agreement with numerical computations
for finite clusters and with available experimental data on the $\beta$-Cu$_2$V$_2$O$_2$ compound.
\end{abstract}


\maketitle

\section{Introduction}
\label{sec:1}

In recent years the low-dimensional quantum Heisenberg  antiferromagnets  have been
extensively  studied motivated by experimental research of such systems. In particular,
there is a vast literature devoted to the study of the two-dimensional (2D)
antiferromagnetic Heisenberg model (AFHM) on the square lattice  initiated  by the discovery
of AF long-range order (LRO) in high-temperature  superconductors (for a review
see~\cite{Manousakis91}). According to the Mermin-Wagner theorem~\cite{Mermin66}, in 2D
isotropic Heisenberg magnets quantum spin fluctuations  destroy the magnetic  LRO at finite
temperature. However, it has been shown  that for the spin-$1/2$ 2D  AFHM with
nearest-neighbor (nn) interaction on the square lattice, LRO can occur at zero
temperature, while at finite temperature  only the exponential increase of the AF
correlation length with decreasing temperature is observed~\cite{Chakravarty89}. For the
2D honeycomb lattice with the lower coordination number $z =3$, quantum spin fluctuations are
more intensive and detrimental for the occurrence  of LRO. Studies of the frustrated
Heisenberg model with the AF interaction between the second-nearest ($J_2$) and the third-nearest ($J_3$)
neighbors on the honeycomb lattice have revealed  several phases with LRO, and  a more
complicated  phase diagram occurs.  For instance, using the coupled cluster method
~\cite{Bishop12,Li12}, in the $J_1$ -- $J_2 $ -- $J_3$
Heisenberg model on the 2D honeycomb lattice four  competing magnetic phases
(N\'{e}el, stripe, N\'{e}el-II collinear AF, and spiral phases) were  found
depending on the model parameters. Much attention has been paid to studies of the 2D
Kitaev-Heisenberg model with the isotropic nn interaction $J$ and the bond-depending Kitaev
interaction $K_{\alpha}$~\cite{Kitaev06}. Besides the spin-liquid Kitaev phase at $\, J
=0$,  a rich phase diagram at zero temperature with competing LRO (ferromagnetic,
AF, stripe and zigzag phases) emerges (see, e.g.,~\cite{Chaloupka10,Chaloupka13}). In a model with
anisotropic $K_{\alpha}$ interactions, e.g.,  $K_z > K_x = K_y$, the LRO
survives  at finite  temperature as shown in Ref.~\cite{Vladimirov16}.

For the isotropic 2D  honeycomb  Heisenberg model with nn AF interaction, the  LRO at zero
temperature, similar to the square lattice, was confirmed in a number of studies mostly
performed for finite-lattice systems. The ground-state energy $E_0$, staggered
magnetization $m_{st}$, uniform static susceptibility $\chi(0)$ at $T=0$, and  other
properties of the 2D honeycomb AFHM were calculated using various
methods, such as  extrapolations of finite-lattice diagonalizations and
quantum Monte Carlo (QMC) simulations \cite{Reger89,Low09,Jiang12,Castro06}, spin-wave theory
\cite{Castro06,Weihong91}, series expansions around the Ising limit \cite{Oitmaa92,Oitmaa12},
Schwinger boson method~\cite{Mattsson94}, and variational RVB wave function
approach~\cite{Noorbakhsh09}. In Ref.~\cite{Tsirlin10} it was suggested to consider the
$\beta$-Cu$_2$V$_2$O$_2$ compound  as the best available experimental realization of the spin-$1/2$ AFHM on
the honeycomb-like  lattice. The system can be
characterized by the nearly isotropic nn exchange interaction $J = $(60--66)K and  the
interlayer coupling $J_{z} = 0.2 J$ which results in the N\'{e}el temperature
$T_N \simeq 26$~K.

Besides the  honeycomb AFHM, the honeycomb Hubbard model has been studied too.
In Ref.~\cite{Peres04} the AF LRO was found for a single layer close
to half-filling at large Coulomb repulsion $U$, where the staggered magnetization $m_{st}
= 0.335$ was obtained. In Ref.~\cite{Aria15}, using  the two-particle self-consistent
approach  for the Hubbard model on the honeycomb lattice, the semimetal to spin-liquid
transition was found before the transition to the  AF state.

The AFHM on the honeycomb lattice has been less well studied as compared with the same model
on the square lattice. Most of numerical computations  have been performed for 2D finite-lattice
systems at zero temperature, where such computations for 3D systems  are
difficult to realize. Moreover, the thermodynamics at arbitrary interlayer coupling, e.g., the
dependence of $T_N$ on $J_z$, is not yet developed.
Therefore, analytical approaches which are capable to evaluate the thermodynamics
of the AFHM on the layered honeycomb lattice both in the AF phase and in the
paramagnetic phase with a temperature-dependent AF short-range order (SRO) are
desirable.

To this end, in this paper we present a theory of magnetic order in the honeycomb
AFHM over the whole temperature region that is based on the calculation of
the dynamic spin susceptibility  (DSS) within  the
spin-rotation-invariant (SRI) relaxation-function theory~\cite{Vladimirov05,Vladimirov09}
using the generalized mean-field approximation (GMFA), as has been done in our study of
the compass-Heisenberg model on the square lattice~\cite{Vladimirov15}. Using the result
for the DSS, the staggered magnetization, the static spin susceptibility, the N\'{e}el
temperature as a function of the interlayer coupling, the AF correlation length,
and the spin-excitation dispersion both in
the  AF phase and in the paramagnetic phase are calculated self-consistently, where
we pay  particular
attention to a proper description of AF SRO.
It should be pointed out that the GMFA has been successfully applied to
several quantum spin systems (see, e.g., Refs.~\cite{KY72,ST91,BB94,WI97,Ihle99,Siurakshina00,Siurakshina01,ISW99,YF00,BCL02,Schmalfuss06,HRI08,JIB08,Junger09,MKB09,BMS11,HRG13,Schmalfuss04,Schmalfuss05}). In
particular, let us mention an application of the GMFA to study the spin-$1/2$ AFHM
on the stacked kagom\'{e} lattice~\cite{Schmalfuss04}, where no LRO was found even
for a strong interlayer coupling due to the frustration character of the AF interaction on
this  lattice.

In Sect. \ref{sec:2} we formulate the model and give equations for the DSS. The solution of the
self-consistency equations for the spin correlation functions and the spin-excitation  spectrum in
the GMFA is presented in Sect. \ref{sec:3}. The numerical results and discussion are given in
Sect. \ref{sec:4}. The conclusion can be found in Sect. \ref{sec:5}.

\section{Model and dynamic spin susceptibility}
\label{sec:2}
\begin{figure}
\resizebox{0.42\textwidth}{!}{\includegraphics{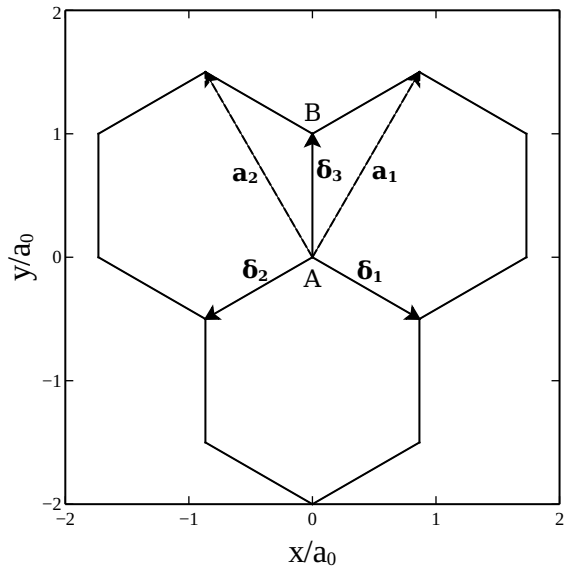}}
\caption[]{Sketch of one honeycomb layer in the congruently stacked lattice, where ${{\bf \delta}_1},\,
{{\bf \delta}_2},\, {{\bf \delta}_3} $ are the intralayer nearest-neighbor
vectors (\ref{nn}), and ${\bf a}_1$, ${\bf a}_2$ are the lattice vectors. }
 \label{fig1}
\end{figure}

We consider congruently stacked honeycomb layers shown in Fig.~\ref{fig1}.
The lattice is bipartite with two triangular sublattices $A$ and $B$. Each site on the
$A$ sublattice is connected to three nn sites belonging to the $B$ sublattice by vectors
${{\bf \delta}_j}$, and sites on $B$ are connected to $A$ by vectors
$-{{\bf \delta}_j}$:
\begin{equation}
{{\bf \delta}_1} = \frac{a_0}{2}(\sqrt{3}, -1),\; {{\bf \delta}_2} = -
\frac{a_0}{2}(\sqrt{3}, 1), \; {{\bf \delta}_3} = a_0(0, 1). \label{nn}
\end{equation}
The basis vectors in the layer  are ${\bf a}_1 = {{\bf \delta}_3} -
{{\bf \delta}_2} = ({a_0}/{2})(\sqrt{3}, 3)$ and ${\bf a}_2 =
{{\bf \delta}_3} - {{\bf \delta}_1} = ({a_0}/{2})(-\sqrt{3}, 3)$, the
lattice constant is $a = |{\bf a}_1| = |{\bf a}_2| = \sqrt{3}a_0$, where $a_0$ is  the nn
distance (see  Fig.~\ref{fig1}); hereafter we put $a_0 = 1$. The reciprocal lattice in the layer is defined by the
vectors ${\bf k}_1 = ({2\pi}/{3}) (\sqrt{3}, 1)$  and $ {\bf k}_2 =({2\pi}/{3})
(-\sqrt{3}, 1)$.

The Heisenberg model on this layered honeycomb lattice can be written as
\begin{equation}
H = \frac{J}{2}\sum_{\langle i\alpha, j\beta \rangle_{xy}}\; {\bf S}_{i\alpha} \; {\bf S}_{j\beta}  +
\frac{J_z}{2}\sum_{\langle i\alpha, j\alpha \rangle_{z}} {\bf S}_{i\alpha} \; {\bf S}_{j\alpha},
 \label{b1}
\end{equation}
where $i$, $j$ count the unit cells, $\alpha$ and $\beta$ are the sublattice indexes,
$\langle i\alpha, j\beta \rangle_{xy}$ and $\langle i\alpha, j\beta \rangle_{z}$
denote nn sites in the ${xy}$ plane and along the $z$ direction, respectively,
$J > 0$ is the  AF intralayer exchange interaction, and $J_z$ is the coupling between the layers.

To calculate the spin-excitation spectrum  and to evaluate the thermodynamic quantities in the
model  (\ref{b1}), we consider the    retarded two-time commutator Green function
(GF)~\cite{Zubarev60}:
\begin{eqnarray}
 {G}^\nu_{ij,\alpha\beta} (t-t')& = &  -i \theta(t-t') \langle [{S}^\nu _{i \alpha}(t) \;
   , \; {S}^\nu _{j\beta}(t') ]\rangle
\nonumber\\
 & \equiv & \langle \!\langle {S}^\nu_{i \alpha}(t) \mid {S}^\nu  _{j \beta}(t')\rangle \!\rangle,
     \label{b2}
\end{eqnarray}
where $ [A, B] = AB - BA$,  $\, A(t)= \exp (i Ht) A\exp (-i Ht)$, and
 $\,\theta(x)$ is the Heaviside function. In the SRI theory all GF components $\nu = x, y, z$ are
 equivalent; therefore, we consider only one of them, e.g., $\nu = x$.  The Fourier representation in $({\bf q}, \omega) $-space is defined by the relation:
\begin{eqnarray}
 {G}^\nu_{ij,\alpha\beta} (t-t') &=&
\int_{-\infty}^{\infty}\frac{d\omega }{2\pi} e^{- i\omega(t-t')}
\nonumber\\
&\times&\frac{1}{N}\,
 \sum_{\bf q}{\rm e}^{i{\bf q ({\bf R}_i-{\bf R}_j)}} {G}^\nu _{\alpha\beta}({\bf q},\omega) ,
     \label{b3}
\end{eqnarray}
where $N$ and ${\bf R}_i$ is the number and position of unit cells, respectively, and the GF matrix reads
\begin{eqnarray}
&& {G}^\nu({\bf q},\omega) = \left(
\begin{array}{cccc}
  \langle \! \langle  S^\nu_{{\bf q} A} | S^\nu_{{\bf q} A} \rangle\! \rangle_\omega  \;
  \langle \! \langle  S^\nu_{{\bf q} A} | S^\nu_{{\bf q} B} \rangle\! \rangle_\omega  \\
  \langle \! \langle  S^\nu_{{\bf q} B} | S^\nu_{{\bf q} A} \rangle\! \rangle_\omega  \;
  \langle \! \langle  S^\nu_{{\bf q} B} | S^\nu_{{\bf q} B} \rangle\! \rangle_\omega  \\
\end{array}\right) .
\label{b3a}
\end{eqnarray}
In the relaxation-function theory developed on the basis of the
equation of motion method in
Refs.~\cite{Vladimirov05,Vladimirov09} we obtain the following representation of the
DSS $\chi({\bf q},\omega) = -{G}^\nu({\bf q},\omega)$:
\begin{equation}
\chi ({\bf q},\omega) =
[ F({\bf q}) + \Sigma({\bf q},\omega)  - \omega^2
\, \tau_0 \, ]^{-1}\times m ({\bf q}).
     \label{b4}
\end{equation}
Here, $F({\bf q})$  is the  frequency matrix of spin excitations in the GMFA,
where the approximation
$-\ddot{S}^{\nu}_{{\bf q}\alpha}= [[S^{\nu}_{\bf q \alpha}, H], H]=\Sigma_{\beta} F_{\alpha \beta}({\bf q}) S^{\nu}_{{\bf q}\beta}$
is made, $\tau_0$  is the unity matrix, and $m ({\bf q})$ is the moment matrix with components
 $m_{\alpha \beta}({\bf q}) = \langle [i \dot{S}^{\nu}_{{\bf q}\alpha} ,
S^{\nu}_{{\bf q}\beta}] \rangle = \langle [[S^{\nu}_{{\bf q}\alpha}, H], S^{\nu}_{{\bf
q}\beta}] \rangle$. The self-energy $\Sigma({\bf q},\omega)$ can be expressed exactly by a multispin GF
(see Refs.~\cite{Vladimirov05,Vladimirov09}).

\section{Generalized  mean-field approximation}

\label{sec:3}

We consider the GMFA  for the DSS neglecting the self-energy
$\Sigma({\bf q},\omega)$ in Eq.~(\ref {b4}). Then, for a
lattice with basis the zero-order DSS is given by ~\cite{Schmalfuss04}:
\begin{equation}
[F({\bf q})  - \omega^2 \, \tau_0 \, ] \times
  \chi ({\bf q},\omega) =  m({\bf q}).
     \label{b6}
\end{equation}
For the static spin susceptibility we obtain
 \begin{equation}
\chi_({\bf q}, 0) \equiv \chi_({\bf q}) =
F^{-1}({\bf q}) \times m ({\bf q}).
 \label{b6a}
\end{equation}
The direct calculation  of the matrix elements $m_{\alpha \beta}({\bf q})$ yields
\begin{equation}
m ({\bf q}) = \left(  \begin{array}{cc} m_{AA}({\bf q}) & m_{AB}({\bf q})\\
  m_{AB}^*({\bf q}) & m_{AA}({\bf q}) \end{array}  \right),
 \label{b7}
 \end{equation}
where
\begin{eqnarray}
&&m_{AA}({\bf q}) = -6 J C_1 - 4 J_{z} C_{z} (1-\cos q_z), \nonumber\\
&&m_{AB}({\bf q}) = 2  J C_1 \gamma_1({\bf q}) , \quad
\gamma_1({\bf q}) = \sum_i \exp (i {\bf q} {{\bf \delta}_i}), \nonumber\\
&&|\gamma_1(q)|^2 = {1 + 4 \cos (\frac{\sqrt{3}}{2} q_x)[\cos (\frac{\sqrt{3}}{2} q_x) + \cos (\frac{3}{2} q_y)]}.\nonumber
\end{eqnarray}
From the symmetry of our model it is obvious that we have only two
different nn correlation functions: $C_1$ within the plane and $C_z$ between neighboring planes.

To calculate  the frequency matrix $F({\bf q})$ in Eq.~(\ref
{b6}), we start from the second derivative $-\ddot{S}^{\nu}_{i\alpha}$ that is
proportional to  products of three spin operators  on different lattice sites along
nn sequences, e.g., $\langle i A, j B, k A \rangle $. We perform the decoupling of them
as follows
\begin{eqnarray}
S^x_{iA} S^y_{jB} S^y_{kA} &  = & \alpha_1 \langle S^y_{jB} S^y_{kA} \rangle S^x_{iA} =
\alpha_1  C_{1}\,S^x_{iA},
\label{b12a}\\
S^x_{jB}S^y_{iA} S^y_{kA} & = & \alpha_2 \langle S^y_{iA} S^y_{kA} \rangle S^x_{jB}
 = \alpha_2 C_{2} \, S^x_{jB},
 \label{b12b}\\
S^x_{jB}S^y_{iA} S^y_{i+i_z A} & = & \alpha_z \langle S^y_{iA} S^y_{i+i_z A} \rangle
S^x_{jB}
 = \alpha_z C_{z} \, S^x_{jB},
 \label{b12c}
\end{eqnarray}
where the index $i+i_z$ denotes the unit cell with position ${\bf R}_{i+i_z} = {\bf R}_i + {\bf a}_3$.
Here the vertex renormalization parameters $\alpha_1$ and $\alpha_2$ are attached to the
nn and the next-nn  correlation functions $C_1$ and $C_2$ respectively.
In the 3D case we introduce $\alpha_z$ associated with the nn interlayer
correlation function $C_z$.
For the next-nn correlation functions between the layers $C_{zz} = \langle S^y_{iA} S^y_{i+2i_z A} \rangle$ and
$C_{1z} = \langle S^y_{iA} S^y_{j+i_z B} \rangle$ we
attach the same vertex parameter $\alpha_2$ as for the next-nn within the layer. Using
these decouplings we obtain the frequency matrix $F({\bf q})$:
\begin{equation}
F({\bf q}) = \frac{1}{2}\left(  \begin{array}{cc} F_{AA}({\bf q}) &
F_{AB}({\bf q})
\\ F^*_{AB}({\bf q}) & F_{AA}({\bf q}) \end{array} \right) ,
\label{b13}
\end{equation}
where
\begin{eqnarray}
F_{AA}({\bf q}) & = &  J^2 (3 + 24 \alpha_2 C_2 + 4 \gamma_2({\bf q}) \alpha_1 C_1)
\nonumber \\
 & + &  24 J J_{z} [\alpha_2 C_{1z}(2-\cos q_z) - \cos q_z \alpha_1 C_{1}]\nonumber\\
 & + & J^2_{z} (2 + 8 \alpha_2 C_{zz}) (1-\cos q_z),
 \nonumber \\
F_{AB}({\bf q}) & = & -  \gamma_1({\bf q})\, f_{AB}({\bf q}), \nonumber\\
f_{AB}({\bf q}) & = & 8 J J_{z} [\alpha_{z} C_{z} (1 - \cos q_z) + \alpha_2 C_{1z}
- \cos q_z \alpha_1 C_{1}]
\nonumber \\
 & + & J^2 (1 + 8 \alpha_1 C_1 + 8 \alpha_2 C_2), \label{b14}
\end{eqnarray}
with
\begin{eqnarray}
\gamma_2({\bf q}) &=& \sum_{i, j\neq i} \exp (i {\bf q}
({{\bf \delta}_i}-{{\bf \delta}_j})) \nonumber\\
&=& 4 \cos (\frac{\sqrt{3}}{2} q_x) \cos (\frac{3}{2} q_y) + 2 \cos (\sqrt{3} q_x). \label{b15}
\end{eqnarray}

Since the  matrices $m({\bf q})$ and $F({\bf q})$ commute,
it is convenient to solve Eq.~(\ref {b6})
by introducing the eigenvalues $m_{\pm}({\bf q})$ and the normalized eigenvectors
$|E_{\pm}({\bf q})\rangle$ of the matrix~(\ref {b7}):
\begin{equation}
[m({\bf q}) - \tau_0 m_{\pm}({\bf q})] |E_{\pm}({\bf q})\rangle = 0 , \label{b10}
\end{equation}
which are given by
\begin{eqnarray}
m_{\pm}({\bf q})& = & - 2 J C_1 (3 \pm |\gamma_1(q)|)
 - 4 J_{z} C_{z} (1-\cos q_z),
\nonumber\\
|E_{\pm}({\bf q})\rangle & =& \frac{1}{\sqrt{2}}\left(  \begin{array}{cc} \mp \gamma_1
({\bf q}) / |\gamma_1 ({\bf q})|\\ 1 \end{array}  \right). \label{b11}
\end{eqnarray}
For the same eigenvectors  $|E_{\pm}({\bf q})\rangle$  the spin-excitation
frequencies are obtained as
\begin{equation}
\omega^2_{\pm}({\bf q}) = \frac{1}{2}(F_{AA}({\bf q}) \pm |  \gamma_1({\bf q})
|f_{AB}({\bf q})).
 \label{b17}
\end{equation}
In this notation the DSS reads:
\begin{equation}
\chi_{\alpha \beta} ({\bf q},\omega) = \sum_{j = \pm} \chi_{j} ({\bf q}, \omega)
 \langle \alpha | E_j({\bf q}) \rangle \langle
E_j({\bf q}) | \beta \rangle,
 \label{b16}
\end{equation}
where
\begin{equation}
\chi_{\pm} ({\bf q}, \omega) = \frac{m_{\pm}( {\bf q})} {\omega_{\pm}^2({\bf q}) - \omega^2},
\label{chi_pm}
\end{equation}
and $\langle \alpha | E_{\pm} \rangle\langle E_{\pm} | \beta\rangle = 1/2$ for $\alpha
= \beta$,  otherwise $\langle \alpha | E_{\pm} \rangle\langle E_{\pm} | \beta\rangle =
\mp \gamma_1({\bf q}) / (2|\gamma_1({\bf q})|)$.

Introducing the Fourier representation for the correlation function as in Eq.~(\ref
{b3}),
 \begin{eqnarray}
 C_{ij,\alpha\beta} = \langle{S}^\nu_{i\alpha} S^{\nu}_{j\beta}\rangle =
\frac{1}{N}\, \sum_{\bf q}{\rm e}^{i{\bf q ({\bf R}_i-{\bf R}_j)}} C _{\alpha\beta}(\bf q),
     \label{b18}
\end{eqnarray}
and using the spectral representation for the GF, we calculate the spin correlation
function:
\begin{eqnarray}
&&C_{\alpha \beta}({\bf q}) = \langle {S}^{\nu}_{{\bf q}\alpha} {S}^{\nu}_{{\bf q}\beta}
\rangle \nonumber\\
&&= \sum_{j = \pm} \frac{m_{j} ({\bf q})}{2 \omega_{j}({\bf q})} \coth
\frac{\omega_{j}({\bf q})} { 2 T}\langle \alpha | E_j({\bf q}) \rangle \langle E_j({\bf
q})| \beta \rangle.
 \label{b19}
\end{eqnarray}
 The correlation
functions $C_{\bf R}$ are written as:
\begin{eqnarray}
&&C_{{\bf R}, \alpha \beta} = \langle {S}^{\nu}_{0 \alpha} {S}^{\nu}_{{\bf R}, \beta}
\rangle \nonumber\\
&&= \frac{1}{N}\sum_{{\bf q} \neq {\bf Q}} C_{\alpha \beta}({\bf q}) e^{i \bf qR} +
C_{\alpha \beta} e^{i \bf QR}, \label{b16a}
\end{eqnarray}
where $C_{\alpha\alpha} = - C_{\alpha \neq \beta} =  C$,
and  the wave-vector ${\bf Q}$
characterizes the LRO.  The condensation part $C$ appears in the ordered phase when
$\omega_+({\bf q})$ condensates at ${\bf Q}$  which determines the LRO,
$\omega_+({\bf Q}) = 0$.
In the  case of AF  order in the two-sublattice model we have ${\bf Q} =
(0, 0, \pi)$, and the staggered magnetization $m_{st}$ is determined  by
\begin{equation}
(m_{st})^2 = 3 \, C.
 \label{b17a}
\end{equation}
Let us consider the uniform static susceptibility
$\chi = \chi_{AA}(0) + \chi_{AB}(0) = \chi_-(0)$ and the staggered susceptibility
$\chi_{st} = \chi_{AA}({\bf Q}) - \chi_{AB}({\bf Q}) = \chi_+({\bf Q})$.
Expanding $\chi_-({\bf q})$ around ${\bf q} = 0$ we get
\begin{equation}
 \chi = \frac{- 4 C_1}{J + 2 J \alpha_1 \,C_1 +
 8 J \alpha_2 \,C_2 + 8 J_z (\alpha_2 C_{1z} - \alpha_1 C_{1})}.
 \label{b19a}
\end{equation}
Considering ${\bf q} \Rightarrow 0$ from different directions in momentum space we can
write  the  isotropy condition for the uniform static spin susceptibility
which  should not depend on the direction of $q \Rightarrow 0$: $\lim_{q_{x,y} \rightarrow 0}
\chi_-({\bf q})|_{q_z=0} =
 \lim_{q_z \rightarrow 0} \chi_-({\bf q})|_{q_{x,y}=0} $
(see Refs. \cite{Ihle99,Siurakshina00,Schmalfuss04,Schmalfuss06,Junger09}). This
condition gives us  the relation between  the intralayer and interlayer correlation
functions:
\begin{eqnarray}
&&\frac{- 4 C_1}{1 + 2  \alpha_1 \,C_1 + 8 \alpha_2 \,C_2 + 8 (J_z/J)\, (\alpha_2 C_{1z} -
\alpha_1 C_{1})}\nonumber\\
&&= \frac{- 2 C_z}{12  (\alpha_2 C_{1z} - \alpha_{z} C_{z}) + (J_z/J)\,
 (1 + 4 \alpha_2 C_{zz})}.
 \label{b20}
\end{eqnarray}
We expand $\chi_+({\bf q})$ in the neighborhood of the AF vector ${\bf Q}$ and obtain
$\chi_+({\bf Q} + {\bf k}) = \chi_+({\bf Q})[1+\xi_{xy}^2 (k_x^2 + k_y^2) + \xi_z^2 k_z^2 ]^{-1}$,
where for the intralayer correlation length $\xi_{xy}$ we get:
\begin{eqnarray}
&&\xi_{xy}^2 = - \frac{1}{2\omega_+^2({\bf Q})} [J^2(3  + 42  \alpha_1 C_1 + 24  \alpha_2 C_{2}) \nonumber\\
&&+ 24 (JJ_z)\, (2 \alpha_{z} C_{z} + \alpha_2 C_{1z} + \alpha_1 C_{1})].
 \label{b20a}
\end{eqnarray}
At zero temperature in the 2D case or at $T = T_N$ in the 3D case, the LRO occurs when both the
correlation length (\ref{b20a}) and $\chi_+({\bf Q})$ diverge.

\section{Results}
\label{sec:4}
\begin{table}
\caption{ Results for the 2D honeycomb AF Heisenberg model at $T
= 0 $: ground-state energy per site $E_0/N_s$, staggered magnetization $m_{st}$ and uniform
static susceptibility $\chi(0)$ at $T=0$, obtained by QMC~\cite{Reger89,Low09,Jiang12},
spin-wave theory~\cite{Weihong91}, series expansion \cite{Oitmaa92,Oitmaa12}, slave boson
method~\cite{Mattsson94}, variational RVB wave function~\cite{Noorbakhsh09}, coupled
cluster method~\cite{Bishop12}, and by our theory using $m_{st}$ as input.}
 \label{Table1}
\begin{tabular}{crrc}\\
\hline
   Refs  &  $\; E_0 /J N_s \; $ & $ \; m_{st}\quad  $   & $ \; \chi(0)\, J \; $  \\
\hline
\cite{Reger89} & $\;$ - 0.5445 $\;$ &  0.22(3) $\;$ & $\;$  - $\;$    \\
\cite{Low09} & $\;$  - 0.5445 $\;$ & 0.2681 $\;$  & 0.05188 $\;$ \\
\cite{Jiang12}  & $\;$ - $\;$ &  0.2688 $\;$ &  -$\;$ \\
\cite{Weihong91} & $\;$ -0.5485 $\;$ &  0.2418 $\;$  & 0.0457 $\;$ \\
\cite{Oitmaa92}  & $\;$ -0.5443 $\;$ & 0.266(9) $\;$  & 0.0756 $\;$  \\
\cite{Oitmaa12} & $\;$ -0.54 $\;$ & 0.265 $\;$  & - $\;$  \\
\cite{Mattsson94} & $\;$ - $\;$ & -  $\;$  & 0.0683 $\;$  \\
\cite{Noorbakhsh09}  & $\;$ -0.5440 $\;$ & 0.26 $\;$  & - $\;$  \\
\cite{Bishop12} & $\;$ -0.53 $\;$ & 0.272 $\;$  & - $\;$ \\
Our results & $\;$ - 0.59 $\;$ & 0.2681 $\;$ &  0.056 $\;$   \\
\hline
\end{tabular}
\end{table}

\begin{figure}
\resizebox{0.42\textwidth}{!}{\includegraphics{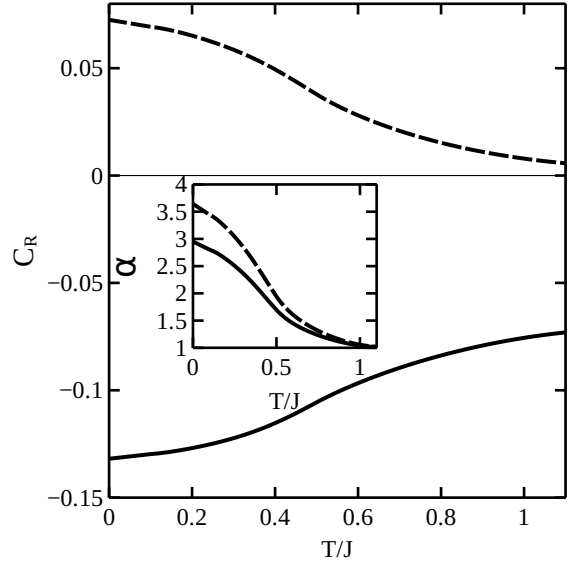}}
\caption[]{Correlation functions $C_1$ (solid line) and $C_2$ (dashed line) at $J_z = 0$ versus
temperature. The inset shows the vertex renormalization parameters $\alpha_1$ (solid) and
$\alpha_2$ (dashed).}
 \label{fig2}
\end{figure}

\begin{figure}
\resizebox{0.42\textwidth}{!}{\includegraphics{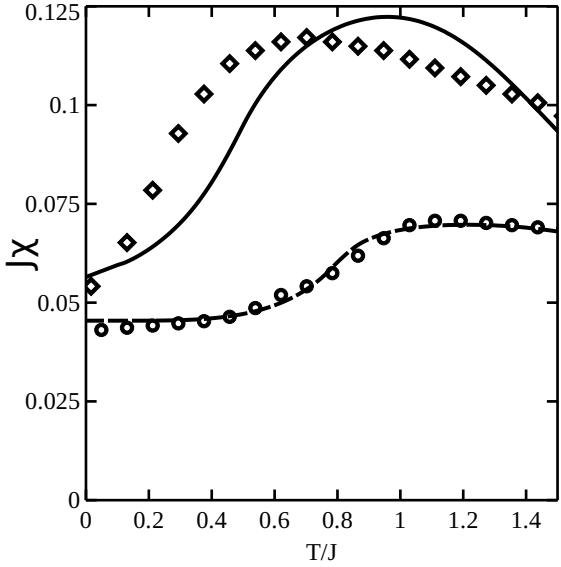}}
\caption[]{Uniform static spin susceptibility $J\chi$ versus
temperature for  $J_z  = 0$ (solid) and  $J_z = J$ (dashed), compared with
the QMC data~\cite{Low09} for $J_z  = 0$ (diamonds) and  $J_z = J$ (circles).}
 \label{fig3}
\end{figure}

\begin{figure}
\resizebox{0.42\textwidth}{!}{\includegraphics{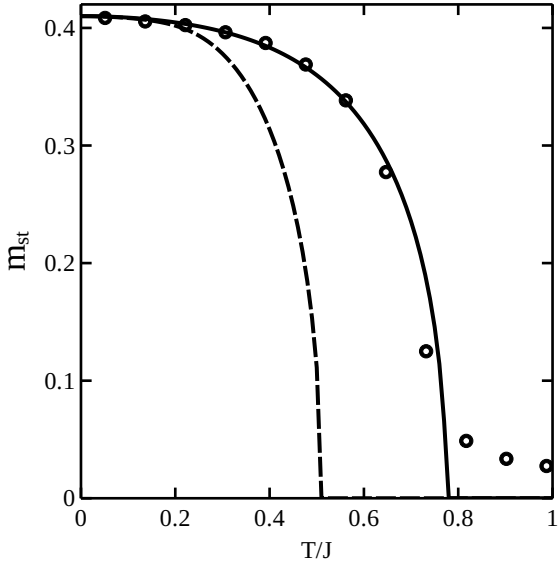}}
\caption[]{Staggered magnetization $m_{st}$ for $J_z  = 0.2J$ (dashed) and $J_z = J$ (solid) versus
temperature compared with the QMC data of Ref.~\cite{Low09} (circles)}
 \label{fig4}
\end{figure}

\begin{figure}
\resizebox{0.42\textwidth}{!}{\includegraphics{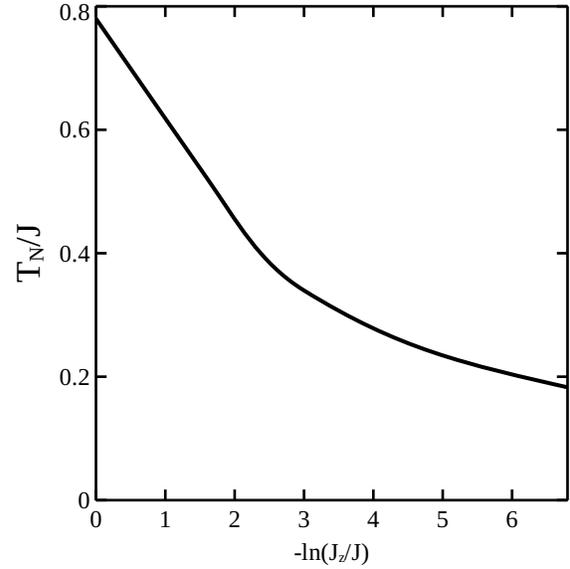}}
\caption[]{N\'{e}el temperature $T_N$ as a function of the interlayer coupling $J_z$.}
 \label{fig5}
\end{figure}

\begin{figure}
\resizebox{0.42\textwidth}{!}{\includegraphics{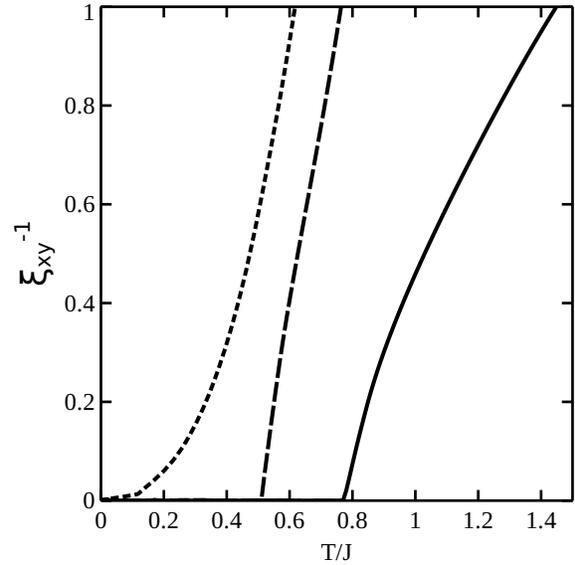}}
\caption[]{Inverse intralayer correlation length $\xi_{xy}^{-1}$ versus temperature for the interlayer coupling $J_z
= 0$ (dotted), $J_z = 0.2J$ (dashed), $J_z = J$ (solid).}
 \label{fig6}
\end{figure}

\begin{figure}
\resizebox{0.42\textwidth}{!}{\includegraphics{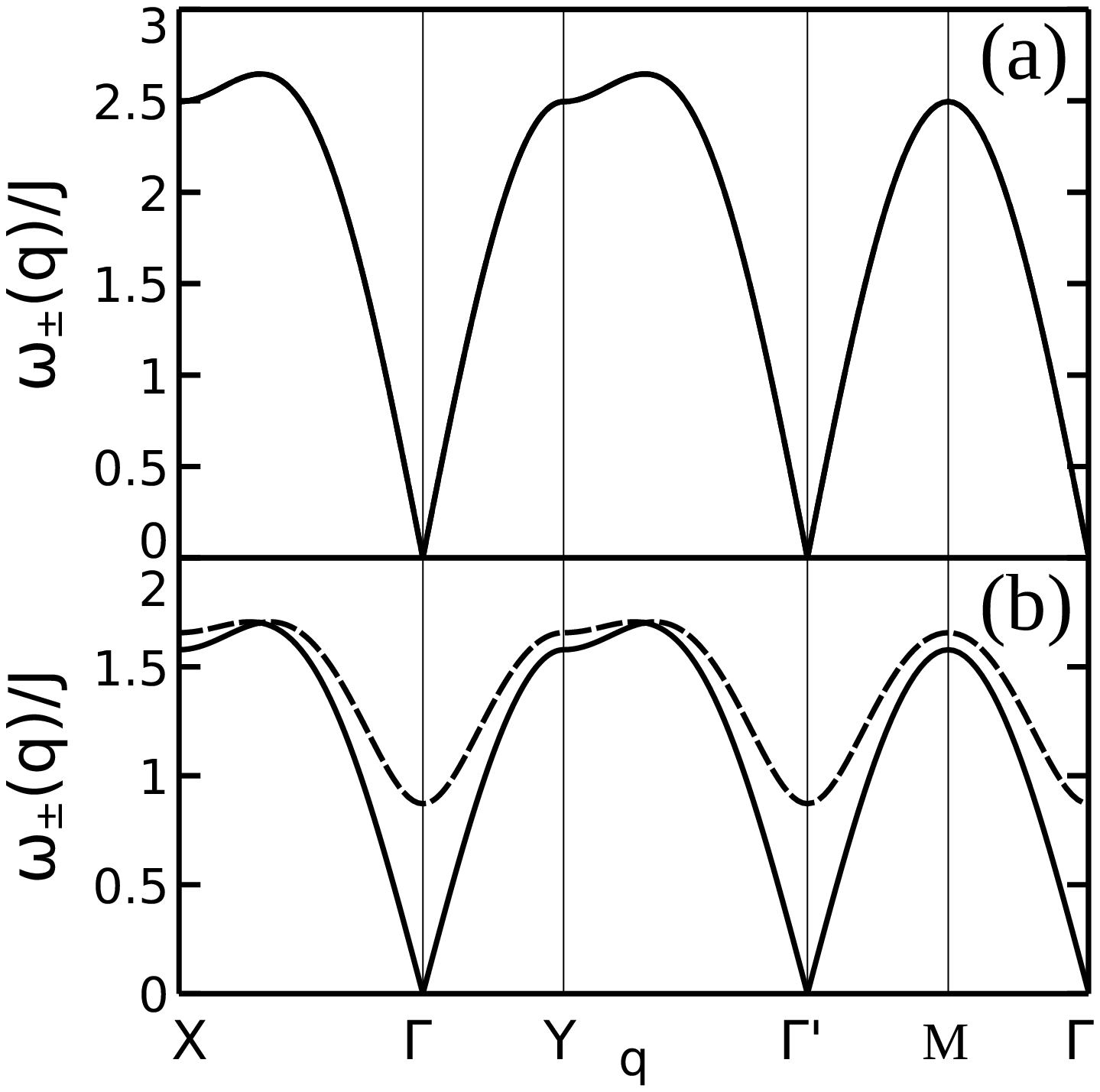}}
\caption[]{Spin-excitation spectrum at $J_z = 0$: $\omega_-({\bf q})$ (solid)
and $\omega_+({\bf q})$ (dashed) for (a) $T = 0$ and (b) $T = 0.6J$.}
 \label{fig7}
\end{figure}

\begin{figure}
\resizebox{0.42\textwidth}{!}{\includegraphics{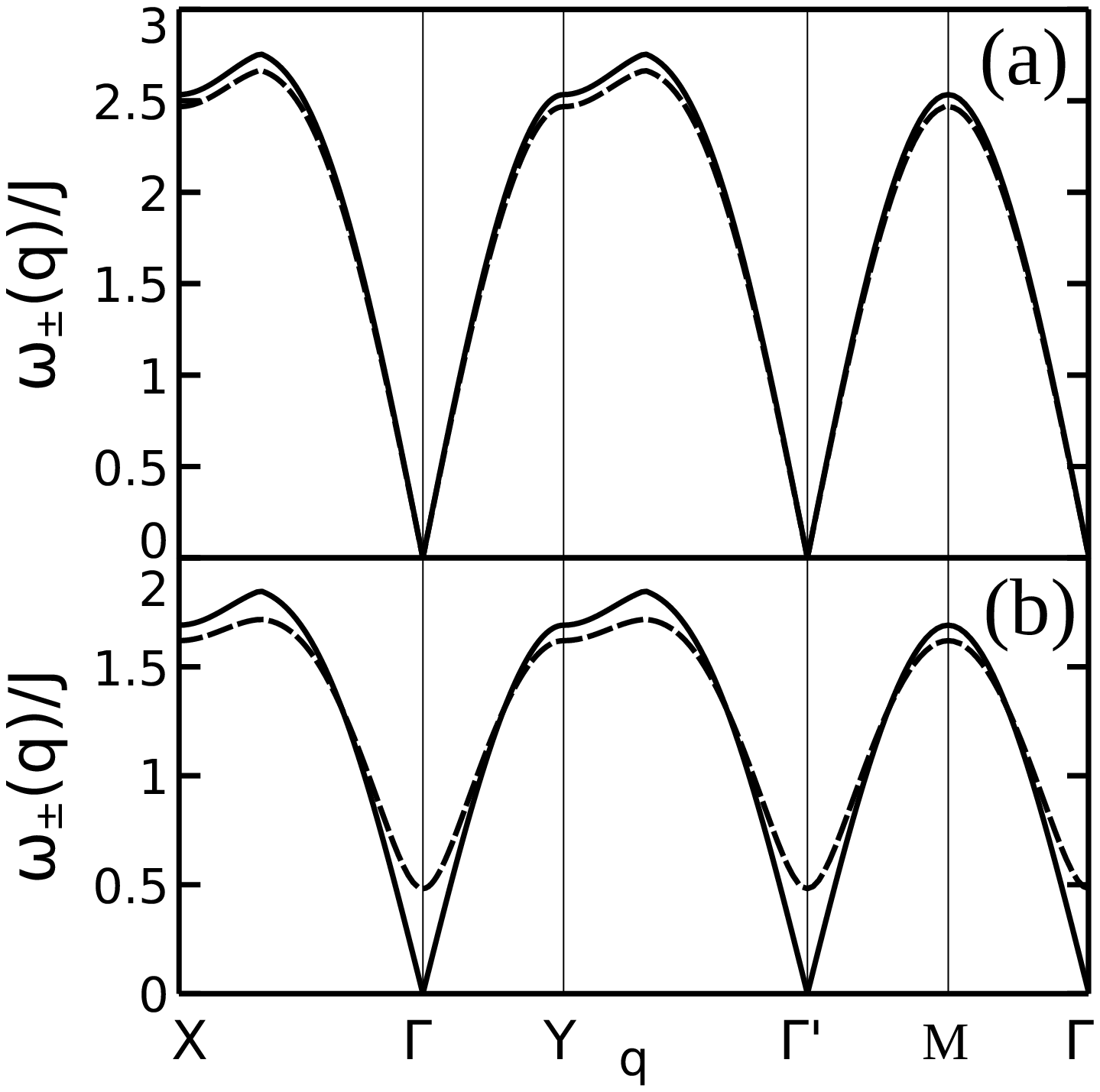}}
\caption[]{Spin-excitation spectrum at $J_z = 0.2J$: $\omega_-(q_z = 0)$ (solid)
and $\omega_+(q_z = \pi)$ (dashed) for (a) $T = 0$ and (b) $T = 0.6J$.}
 \label{fig8}
\end{figure}

To evaluate the spin-excitation spectrum and the thermodynamic properties, the correlation
functions $C_{{\bf R}, \alpha \beta}$ and the vertex parameters $\alpha_1$, $\alpha_2$, and
$\alpha_z$ appearing in the spectrum $\omega_{\pm}({\bf q})$ as well as the
condensation term $C$ in the LRO phase have to be determined as solution of a
coupled system of self-consistency equations. Besides Eq.~(\ref{b16a})
for calculating the correlation functions, we have the sum rule $C_{{\bf R} = 0, \alpha
\alpha} =  \langle {S}^{\nu}_{i\alpha} {S}^{\nu}_{i\alpha} \rangle = 1/4\; $,
the isotropy condition (\ref{b20}), and the LRO condition $\omega_{+}({\bf Q}) = 0$.
That is, we have more parameters than equations. To obtain a closed system of
equations, we adjust the staggered magnetization at $T=0$
to the QMC values $m_{st}(0) = 0.2681$ for $J_z = 0$ and $m_{st}(0) = 0.41$ for $J_z = J$~\cite{Low09},
where we take the latter value in the region $0 < J_z \leq 1$.
For finite temperatures we follow Refs.~\cite{ST91,WI97,Siurakshina00,Siurakshina01,Junger09} and use
the ansatz
$r_{\alpha}(T) \equiv [\alpha_2(T) - 1]/[\alpha_1(T) - 1] = r_{\alpha}(0)$.

The results of our self-consistent calculations are presented in Table~\ref{Table1}
 and in Figs.~\ref{fig2}  - \ref{fig8}.
 As can be seen from Table~\ref{Table1}, our
results for the ground-state energy $E_0/ J N_s$ per site and for the
zero-temperature uniform static susceptibility $\chi(0)$
rather well agree with the calculations by various methods.
Note that the ground-state energy per site taken from Ref.~\cite{Low09},
$E_0/JN_s=-0.5445$, is related to the ground-state energy $e_0/J$ per bond, $e_0/J=-0.36303$ obtained
in Ref.~\cite{Low09}, by $E_0/N_s = \frac{3}{2}e_0$.
In Fig.~\ref{fig2} the nn and next-nn correlation functions in the 2D model are plotted.
With increasing temperature the AF SRO, reflected in alternating signs of $C_1$ and $C_2$,
is weakened which corresponds to a decrease of the correlation
functions and vertex parameters (see inset of Fig.~\ref{fig2}).

The uniform static spin susceptibility $\chi$ is shown in Fig.~\ref{fig3} for the 2D and
3D ($J_z = J$) cases.
The increase of $\chi$ with temperature is due to the reduction of AF short-range
correlations which is connected with a weakening of the spin stiffness against the
orientation of spins along a homogeneous magnetic field. The further decrease of
AF SRO results in a crossover to the high-temperature Curie-Weiss behavior, so that
$\chi$ reveals a maximum at a temperature of the order of the exchange interaction $J$.
For $J_z = 0$ we obtain the maximum value $\chi^{max} = 0.122/J$ at $T^{max} = 0.96 J$.
This is close to the QMC result of Ref.~\cite{Low09}: $\chi^{max} = 0.117/J$ at $T^{max} = 0.72 J$.
In the 3D case with $J_z = J $ our results for the temperature
dependence of $\chi$ in Fig.~\ref{fig3} are in a very good agreement with the
QMC calculations in Ref.~\cite{Low09}. The interlayer coupling $J_z > 0$
lowers the susceptibility and slightly shifts the maximum of $\chi$ to higher
temperatures in comparison with the 2D case, which is due to the $J_z$-induced
enhancement of AF SRO.
Let us compare the maximum temperature $T^{max}$ with the value obtained by susceptibility
experiments on the the $\beta$-Cu$_2$V$_2$O$_2$ compound, $T^{max}_{exp} = 50$K~\cite{He2008}.
Taking $J = 60$K~\cite{Tsirlin10} we get $T^{max} = 58$K which is near to the experimental value.

In Fig.~\ref{fig4} the staggered magnetization $m_{st}$ is depicted. It reveals a second-order
transition to the AF phase below the N\'{e}el temperature $T_N$. For $J_z=0.2J$ we
obtain $T_N = 0.52J$, whereas the QMC simulations of Ref.~\cite{Tsirlin10} yield
the lower value $T_N^{QMC} = 0.41 J$. The inequality $T_N^{GMFA} > T_N^{QMC}$ was also found
for the AFHM on the stacked square lattice~\cite{Junger09}. Considering $\beta$-Cu$_2$V$_2$O$_2$
with $J_z = 0.2J$ and $J=$(60--66)K~\cite{Tsirlin10} we have $T_N \simeq $(30--35)K
that is close to the experimental value $T_N^{exp} = 26$K~\cite{He2008}.
For $J_z = J$ we find
$T_N = 0.78J$ and a good agreement of the temperature dependence
of $m_{st}$ with the QMC data for the largest system size obtained in Ref.~\cite{Low09}.

The N\'{e}el temperature as a function of the interlayer coupling $J_z$ is shown in Fig.~\ref{fig5},
where $T_N$ reveals the strongest decrease with decreasing $J_z$ for $J_z/J \ll 1$ and
tends to zero logarithmically for  $J_z  \rightarrow 0$. The $J_z$ dependence of $T_N$ in the region $0\leq J_z\leq 0.2J$
can be approximated with the accuracy of $1\%$ by the formula
\begin{equation}
\frac{T_N}{J} = \frac{A}{B - \ln (J_z/J)},
 \label{b21}
\end{equation}
where $\,A = 1.48,\, B = 1.25 \,$. This behavior resembles the empirical formula proposed in
Ref.~\cite{Yasuda05} and the result of Refs.~\cite{Junger09,Schmalfuss05}.
Note that qualitatively the same law (\ref{b21}) with $\lim_{J_z \rightarrow 0} \partial T_N / \partial J_z = \infty $ was also found in the random phase approximation
(see Ref.~\cite{Junger09}).

In Fig.~\ref{fig6} the influence of the interlayer coupling on the temperature dependence
of the intralayer correlation length $\xi_{xy}$ is illustrated. In the 2D case,
$\xi_{xy}$ diverges exponentially at $T=0$. For $J_z > 0$, $\xi_{xy}$ diverges
at $T_N$, since the gap $\omega_+({\bf Q})$ closes as $T$ approaches $T_N$ from above.
In the vicinity of $T_N$, $\xi_{xy}^{-1}$ behaves as $T-T_N$ (corresponding to the critical index $\nu=1$),
as it was also found by previous mean-field approaches (see Refs.~\cite{Siurakshina00,Junger09}).

The spectrum of spin excitations $\omega_{\pm}({\bf q})$ is
 shown in Figs.~\ref{fig7} and \ref{fig8}  along the
symmetry directions $X(-1, 0) \rightarrow \Gamma(0, 0) \rightarrow Y(0, 1) \rightarrow
\Gamma'(1, 1)\rightarrow M(1/2,1/2)   \rightarrow \Gamma  $ of the BZ using the
same notation as in Ref.~\cite{Chaloupka13}.
In the LRO phase, i.e., for $T=0$ at $J_z = 0$ and for $T < T_N = 0.52 J$ at $J_z=0.2J$, the
spin excitations are spin waves with gapless branches depicted in Fig.~\ref{fig7}(a) and ~\ref{fig8}(a).
In the paramagnetic phase, spin waves propagating in AF SRO can exist, if their wavelength
is smaller than the correlation length, i.e., if $q > q_c = 2 \pi \xi ^{-1}$.
For temperatures slightly above the transition temperature, where the correlation length is
large enough, this condition can be fulfilled.
The validity region of the spin-wave picture shrinks with increasing temperature, where predominantly
high-energy magnons may be observed. Such a situation was encountered in the study of
dibromo Ni complexes given in Ref.~\cite{Junger05}. For $q < q_c$ the spin-wave picture
breaks down. Considering $T = 0.6 J$ in Figs.~\ref{fig7}(b) and \ref{fig8}(b), for
$J_z = 0 (0.2J)$ we get $\xi^{-1} = 0.86(0.41)$ (see Fig.~\ref{fig6}). Correspondingly, in the
whole $q$ range of Fig.~\ref{fig7}(b) we have $q < q_c$, and the spin excitations may be named ``paramagnon'' excitations
with the energies $\omega_{\pm}({\bf q})$. Note that such excitations have been measured,
for example, by resonant inelastic x-ray scattering on high-T$_c$ superconductors~\cite{Tacon11,Tacon13}. In  Fig.~\ref{fig8}(b), due to the lower value of $q_c$  as compared with  Fig.~\ref{fig7}(b), we have $q$ region
with  $q > q_c $,  where   the spin excitations are high-energy magnons.  Thus, our spin-excitation spectra reveal a smooth crossover from spin-wave to paramagnon behavior depending on the wave number and temperature. In the upper (optical) branch, at $T_N$ a gap
is opening at the $\Gamma$ point, i.e., at the AF wave vector ${\bf Q}$ characterizing the
LRO phase in the two-sublattice model (${\bf Q} = (0, 0)$ at $J_z = 0$ and ${\bf Q} = (0,0,\pi)$
at $J_z > 0$).
As can be seen in Figs.~ \ref{fig7} and \ref{fig8},
the spin-excitation energies are decreasing
with increasing temperature.
Let us point out that the spectrum of spin excitations is calculated in the GMFA which
neglects finite-time effects. To take them into account, one has to determine the self-energy,
as has been done in Refs.~\cite{Vladimirov09}. As it turns out, in the square-lattice
Heisenberg model the damping of spin excitations is quite small. Therefore, we believe
that in our model the inclusion of damping will not qualitatively change our results for
the spin-excitation spectrum.
Provided that a spin-1/2 antiferromagnet with a real honeycomb-lattice structure
may be found, it is desirable to confirm our findings on the optical
branch $\omega_{+}({\bf q})$ by scattering experiments, where the intensity of the
scattering is determined by the
staggered susceptibility $\chi_+ ({\bf Q})$ given by Eq.(\ref{chi_pm}),
so that the gap $\omega_+ ({\bf Q})$ may be measured.

\section{Conclusion}
\label{sec:5}
In this paper we have evaluated thermodynamic quantities and spin-excitation spectra
of the AF Heisenberg model on the stacked honeycomb lattice by calculating the dynamic
spin susceptibility within a spin-rotation-invariant generalized mean-field approach for
arbitrary temperatures. Our main focus was the analysis of the temperature dependence
of the uniform static susceptibility in the paramagnetic phase, which we have explained
in terms of AF SRO, and the calculation of the N\'{e}el temperature in dependence on the
interlayer coupling. Our results are in a good agreement with available QMC
and experimental data. From  this we conclude that our investigation forms a good basis
for forthcoming studies of extended layered honeycomb Heisenberg models
(e.g.,  hole hopping).\\

\acknowledgments
The authors would like to thank J. Richter for valuable
discussions. The financial support by the Heisenberg-Landau program
of JINR is acknowledged.
One of the authors (N. P.) thanks the Directorate of
the MPIPKS for the hospitality extended to him during his stay at the Institute.

\end{document}